# Lattice dynamical analogies and differences between $SrTiO_3$ and $EuTiO_3$ revealed by phonon-dispersion relations and double-well potentials

Jerry L. Bettis[1], Myung-Hwan Whangbo[1], Jürgen Köhler[2], Annette Bussmann-Holder[2], and A. R. Bishop[3]

[1]Department of Chemistry, North Carolina State University, Raleigh, North Carolina 27695-8204, USA

[2]Max-Planck-Institut für Festkörperforschung, Heisenbergstr. 1, D-70569 Stuttgart, Germany

[3]Los Alamos National Laboratory, Los Alamos, NM 87545, USA

**Abstract**

A comparative analysis of the structural phase transitions of $EuTiO_3$ and $SrTiO_3$ (at $T_S$ = 282 and 105 K, respectively) is made on the basis of phonon-dispersion and density functional calculations. The phase transition of $EuTiO_3$ is predicted to arise from the softening of a transverse acoustic zone-boundary mode caused by the rotations of the $TiO_6$ octahedra, as also found for the phase transition of $SrTiO_3$. While the temperature dependence of the soft mode is similar in both compounds, their elastic properties differ drastically due to a large difference in the double-well potentials associated with the soft zone boundary-acoustic mode.

## 1. Introduction

The perovskite oxides SrTiO$_3$ (STO) and EuTiO$_3$ (ETO) are similar in various dynamical properties such as the soft transverse optical modes [1 – 5] as well as the structural [6 – 8] and incipient polar instabilities. The ionic radii of Eu$^{2+}$ and Sr$^{2+}$ are almost identical so that STO and ETO have identical nearly unit cell parameters. Both oxides exhibit incipient ferroelectric instabilities, namely, they are quantum paraelectrics with quantum fluctuations suppressing a polar phase transition [4 – 6]. The ferroelectric transition temperature T$_C$ is estimated to be 37 K for STO, and lower than ~150 K for ETO. The antiferromagnetic (AFM) phase transition of ETO at T$_N$ = 5.5 K [7] has been investigated in detail due to the possibility of multiferroicity in ETO. Besides the strong softening of the long wave length transverse optic (TO) mode, STO exhibits an antiferrodistortive phase transition to a tetragonal phase at T$_S$ = 105 K [8, 9], which is caused by the instability of a transverse acoustic (TA) zone-boundary mode. The symmetry lowering from cubic to tetragonal is accompanied by an extremely small change in the c/a ratio of the lattice parameters [10], which is difficult to detect by conventional diffraction techniques. However, local-probe measurements such as electron paramagnetic resonance (EPR) [8, 11, 12] clearly show the existence of the structural phase transition and demonstrate that the rotation angle of the TiO$_6$ octahedra follows the temperature dependence of an order parameter [13]. In addition, inelastic neutron scattering (INS) experiments have evidenced the softening of the TA mode at the zone boundary [14 – 16]. The soft TO and the soft TA modes have long been considered independent. However, they are related to each other through polarizability effects, which induce a nonlinear coupling between the optic mode and the acoustic mode [17]. An analogous zone-boundary instability has been expected in ETO at higher temperatures because spin lattice coupling is possible and because Eu has a heavier mass than Sr. Indeed, the expected instability of ETO (with samples as prepared in [18]) was confirmed recently by specific heat measurements with $T_S = 282$ K, very close to the calculated $T_S = 298$ K [18]. This newly discovered phase transition unveils an additional analogy between both compounds and could open novel technological applications of ETO and layered superstructures of ETO/STO. Especially, due to the identical valence states and sizes of $Sr^{2+}$ and $Eu^{2+}$ ions, the fabrication of strain-free superstructures would be feasible, and this suggests the possibility of interlayer magnetism, novel multiferroic properties, and even interlayer superconductivity induced by Eu valence instability.

In the present work we examine the origin of the rather large difference in the structural transition temperatures of STO and ETO in some detail by calculating their phonon-dispersion relationships as described in Ref. 18 and by determining the double-well potentials of STO and ETO associated with the antiferrodistortive rotation of their $TiO_6$ octahedra on the basis of first principles density functional calculations. We also examine the spin exchange interactions of ETO by density functional calculations to investigate why it adopts a G-type antiferromagnetic structure below $T_N = 5.3$ K but its dominant spin exchange is ferromagnetic [19, 20].

**2. Phonon Dispersion relations**

As shown in Ref. 18, the temperature dependence of the soft TO mode in ETO is theoretically reproduced in full agreement with experiment by using the *same* lattice dynamical parameters as used for STO. Marked differences between ETO and STO appear in their local double-well potentials, which are exclusively determined by the temperature dependence of the soft mode frequency. The double-well potential of STO is shallow and broad, indicating a displacive behavior, but that of ETO is deep and narrow suggesting a more order-disorder dynamics behavior [18]. The low temperature AFM properties of ETO have been accounted for in terms of an extended polarizability model by adding the coupling of the Eu spins to the $TiO_3$ units and a direct spin-spin interaction term [21]. These latter terms guarantee the coupling of the spins to the soft mode, which is anomalously enhanced at the onset to the AFM order [4, 5].

The interdependence of the zone boundary soft acoustic mode and the long wave length optic mode arises through the nonlinear polarizability of the $TiO_3$ cluster [22 – 25]. This has been calculated self-consistently for each temperature and taken as input to determine the temperature dependence of the zone boundary mode by using the Hamiltonian given in Ref. 19. The results for STO and ETO are compared in Fig. 1.

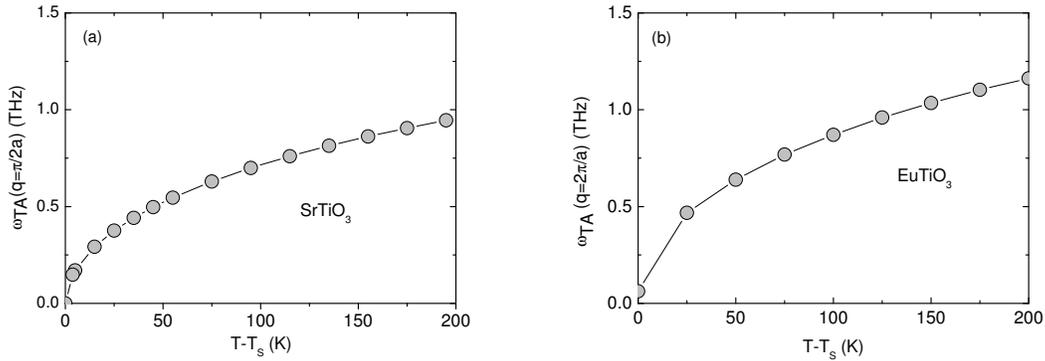

**Figure 1** Temperature dependence of the zone boundary acoustic mode frequency of (a) STO with $T_S$ = 105K and (b) ETO with $T_S$ = 282K,.

It is evident from Fig. 1 that the zone boundary TA modes soften in almost the same manner in STO and ETO. However, by calculating the dispersion of the two lowest lying TA and TO modes of STO and ETO along (110), striking differences between them become apparent as shown in Fig. 2. The TO mode of STO exhibits a strong temperature dependence in the long wave length limit and is very dispersive. However, the TO mode of ETO is almost flat and softens much less with decreasing temperature.

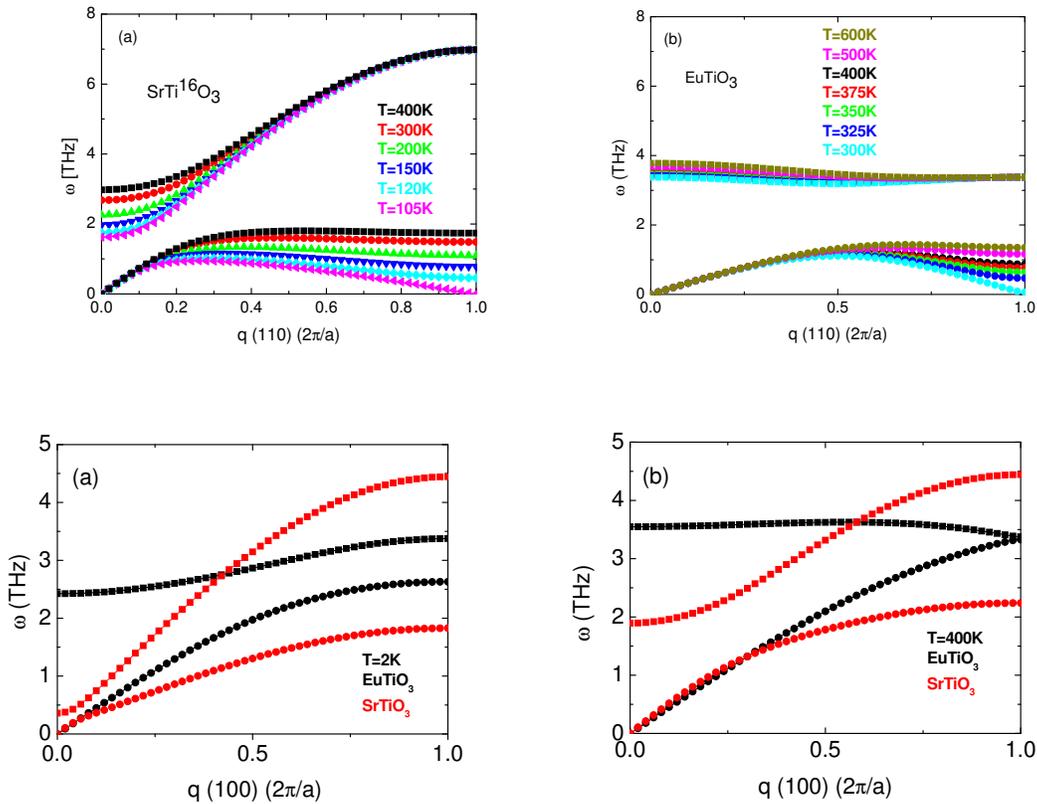

**Figure 2** (Color online) Phonon mode dispersion of the two lowest lying TO and TA frequencies with momentum q along (110) for different temperatures as indicated in the figures for (a) STO, (b) ETO. TA and TO mode dispersion with momentum q along (100) in STO (red) and ETO (black) at (c) T = 2 K and (d) T = 400 K.

In agreement with Fig. 1, the related TA modes have similar zone-boundary temperature dependencies. However, below momentum q ≈ 0.5, characteristic differences appear; the softening in STO continues to the long wave length limit while, in ETO, almost nothing changes with decreasing temperature in this momentum regime. This result indicates that STO has anomalously soft elastic constants, as found experimentally [26 – 28], but ETO does not. These striking differences become even more apparent by normalizing the TA mode frequency to its high temperature value as shown in Fig. 3. In ETO the TA mode is temperature independent up to the momentum q ≈ 0.25. In STO an inflection point is observed at small momentum, which gradually moves to smaller momentum with decreasing temperature. Such an anomalous pre-transitional dispersion evidences fluctuating cluster formation already far above the true instability [13]. Similar findings have been made in BaTiO$_3$ above the ferroelectric phase transition temperature albeit in a less extended temperature regime [29]. The anomalous behavior of the TA mode in STO suggests that the elastic constants exhibit pronounced softening over a wide temperature regime. Since such a softening is absent in ETO, and anomalies in the elastic constants should be absent there.

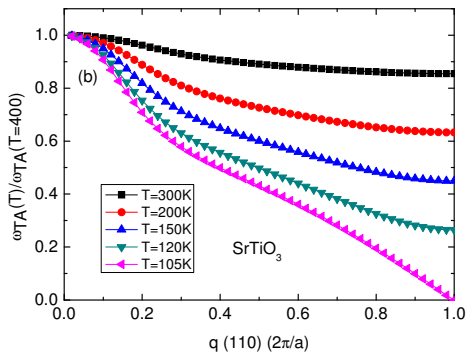 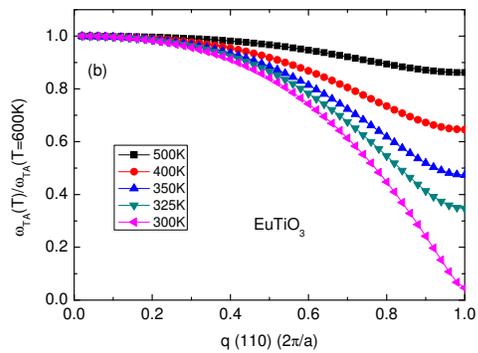

**Figure 3** (Color online) High temperature normalized TA mode dispersion with momentum q along (110) for different temperatures as indicated in the figures for (a) STO and (b) ETO.

By inspecting the dynamical behavior along the (100) direction, the origin of the above described effects becomes evident (Fig. 2c, d). In ETO the TO mode dispersion exhibits an Einstein-oscillator type behavior at 400 K, which is almost unchanged down to 4 K. However, in STO the strongly dispersive TO mode begins coupling with the TA mode at finite momentum already at 400 K, becomes increasingly pronounced at 4 K, and causes strong anomalies in the TA mode dispersion (Figs. 2c, d) [30 – 32]. In ETO, on the other hand, the zone boundary TO and TA modes are degenerate at 400 K due to spin-phonon coupling [18] and a lifting of this degeneracy sets in with decreasing temperature. At T > 400 K, the TA and TO modes exchange their character and a strong zone boundary mode-mode coupling is predicted to take place. For both, ETO and STO, polarizability effects play a role in the TA zone boundary mode also in the (100) direction: some softening is induced with decreasing temperature, although it remains much less pronounced than observed along the (110) direction. Interestingly, the softening of this mode is again comparable in both STO and ETO in spite of the differences in the softening of their TO modes.

## 3. Double-well potentials

To gain insight into the structural zone boundary TA mode instability, we determine the double-well potentials of STO and ETO as a function of the rotation angle θ of the TiO$_6$ octahedra around the c-axis on the basis of first principles density functional calculations, which employ the frozen-core projector augmented wave methods encoded in the Vienna Ab initio Simulation Package (VASP) [33], the generalized-gradient approximation (GGA) [34] with a plane-wave cutoff of 400 eV, and a 6×6×6 k-point mesh for the irreducible Brillouin zone. In the calculations the 298K [35] experimental crystal structure of ETO is utilized without further structure optimization. To describe the effects of electron correlation in the Eu 4f states, the GGA plus on-site repulsion method (GGA+U) [36] is implemented with effective $U_{eff}$ = 4.0 and 6 eV. The Eu$^{2+}$ ion has a half-filled 4f-shell (f$^7$), so that the magnetic insulating state of EuTiO$_3$ is well reproduced by spin-polarized GGA calculations (i.e., GGA+U calculations with $U_{eff}$ = 0), and the double-well potential of EuTiO$_3$ are not expected to be strongly affected by $U_{eff}$ in GGA+U calculations. However, the spin exchange parameters between the Eu$^{2+}$ ions, which are weak in strength, are affected by $U_{eff}$ (see below for further discussion). DFT calculations were performed on a (2a, 2b, 2c) supercell of ETO for ferromagnetic (FM) and AFM G-type spin arrangements. The calculations were constrained to a constant volume at all angles of rotation with the volume given by the experimental one [35]. Only TiO$_6$ rotations within the ab-plane have been considered where nearest neighbor octahedra rotate anti-clockwise with respect to each other (see inset to figure 5). The calculated double-well potentials of STO and ETO are compared in Fig. 4.

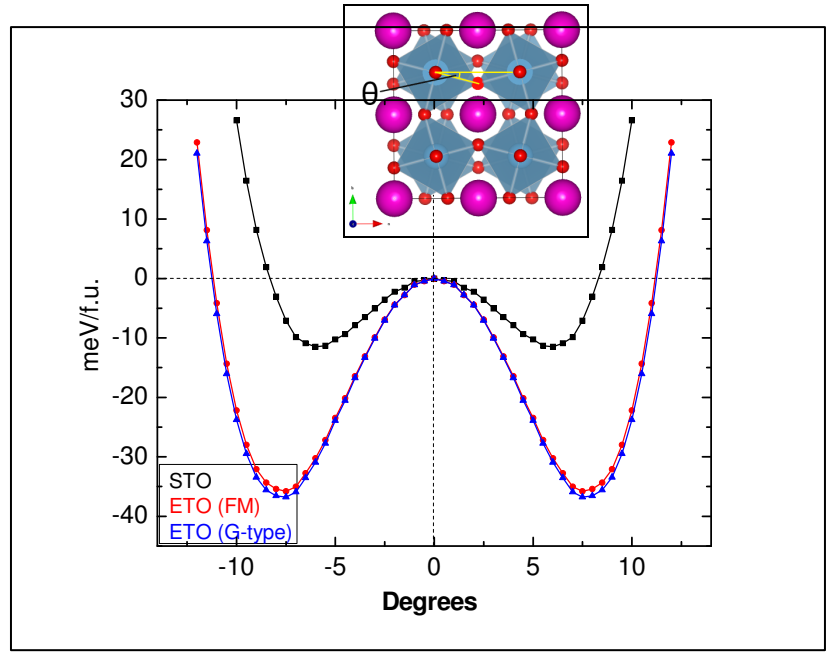

**Figure 4** (Color online) Calculated double-well potential as a function of the oxygen octahedra rotation angle Θ for STO (black), FM ETO (red), and G-Type AFM ETO (blue). The double-well potential shown for EuTiO$_3$ was obtained from the GGA+U calculations with U$_{eff}$ = 4 eV. The inset shows the perovskite structure projected on the ab-plane and the definition of the angle Θ.

For ETO, the double-well potential was evaluated for the above mentioned two different magnetic structures, the G-type AFM and the FM spin states. These two spin states have nearly identical double-well potentials, showing that the effect of magnetic structure on the double-well potential is negligible. In agreement with the previous analysis based on the self-consistent phonon approximation [18], the calculated double-well potential is shallow for STO but very deep for ETO. This massive difference in the double-well potentials indicates that the structural transitions of STO and ETO follow different dynamics, namely, the structural transition of STO is in the displacive limit but that of ETO follows mostly order-disorder dynamics. The displacive character of the transition in STO has been confirmed by various experiments [see, e.g., Ref. 13 and refs. therein]. The predictions for ETO based on our model phonon-dispersion and density functional calculations await experimental verifications. Here the double-well potentials determined from our density functional calculations relate to the zone boundary soft acoustic mode and are a function of the octahedral rotation angle. Thus, it is not straightforward to compare them with those derived for the zone center soft optic mode, which were previously calculated within the self-consistent phonon approximation (SPA) [18]. The latter results provide the double-well potentials as a function of the Ti-O relative displacements. In order to compare both calculations, the rotation angle has to be mapped onto the relative displacement coordinate, which can be performed by a scaling of the double-well potential parameters. A comparison of the corresponding potentials is made in Figure 5.

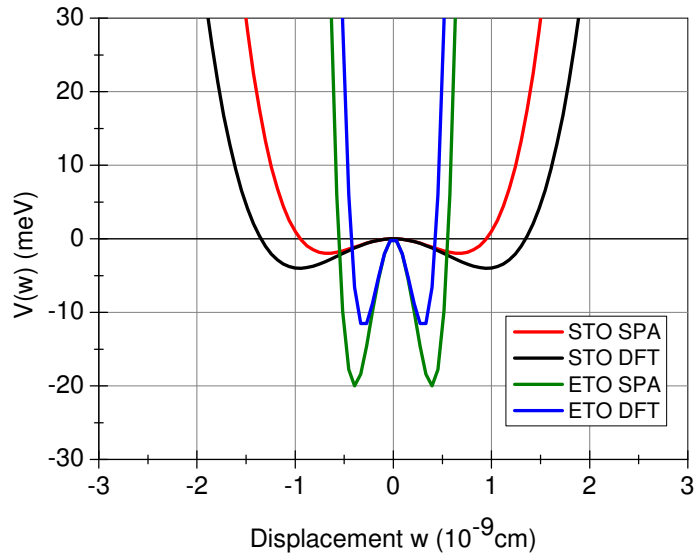

**Figure 5** Comparison of the double-well potentials of ETO and STO as obtained within the SPA and from DFT calculations.

Even though the two sets of the potentials are not exactly in one-to-one correspondence, as might have been expected, the excellent agreement between them is convincing and confirms the above conclusion that the behaviors of ETO and STO are governed by different dynamics.

**4. Spin exchange interaction of ETO**

We evaluate the nearest-neighbor (nn) and next-nearest-neighbor (nnn) spin exchange constants ($J_{nn}$ and $J_{nnn}$, respectively) between $Eu^{2+}$ ($f^7$) ions by performing GGA+U calculations for three ordered spin states constructed by using a (2a, 2b, 2c) supercell of ETO (i.e., the FM, AF1 and AF2 states depicted in Fig. 5). Our calculations employ the all-electron full-potential linearized augmented-plane-wave (FPLAPW) method encoded in the WIEN2k package [37] with $R_{MT}*K_{max}$ = 7.0 and a set of 27 k-points for the irreducible Brillouin zone. The relative energies of the three ordered spin states per formula unit (FU) are listed in Fig. 5. In terms of the spin Hamiltonian

$$\hat{H} = -\sum_{i<j} J_{ij} \hat{S}_i \cdot \hat{S}_j$$

where $J_{ij} = J_{nn}$ or $J_{nnn}$, the total spin-exchange energies, per FU, of the three ordered spin states are obtained as

$$E_{FM} = (-3J_{nn} - 6J_{nnn})(N^2/4)$$
$$E_{AF1} = (-J_{nn} + 2J_{nnn})(N^2/4)$$
$$E_{AF2} = (+3J_{nn} - 6J_{nnn})(N^2/4)$$

by applying the energy expression obtained for spin dimers with N unpaired spins per spin site (in the present case N = 7) [38]. By mapping the relative energies of the three ordered spin states determined by GGA+U calculations onto the corresponding relative energies determined from the above spin-exchange energies, we obtain $J_{nn}/k_B = -0.90$ K and $J_{nnn}/k_B = 0.82$ from the calculations with $U_{eff} = 4$ eV, and $J_{nn}/k_B = -0.04$ K and $J_{nnn}/k_B = 0.64$ K from the calculations with $U_{eff} = 6$ eV. (Note that the AFM spin exchange J depends on the on-site repulsion U as $J \propto -1/U$, so that the AFM J decreases in magnitude with increasing $U_{eff}$ in GGA+U calculations [39].) The G-type AFM state becomes the magnetic ground state, because the spin exchange interactions $J_{nn}$ and $J_{nnn}$ reinforce each other. Each $Eu^{2+}$ ion has six $J_{nn}$ and 12 $J_{nnn}$ interactions so that, in the mean-field approximation [40], the Curie-Weiss temperature θ is related to $J_{nn}$ and $J_{nnn}$ as

$$\theta = \frac{S(S+1)}{3k_B}(6J_{nn} + 12J_{nnn}) = \frac{63}{2k_B}(J_{nn} + 2J_{nnn}),$$

which leads to a positive value in agreement with experiment, i.e., θ = 23.3 and 6.4 K from the $J_{nn}$ and $J_{nnn}$ values calculated with $U_{eff} = 4$ and 6 eV, respectively. Thus, the experimental θ = 5.5 K [35] is well reproduced by the $J_{nn}$ and $J_{nnn}$ values obtained with $U_{eff} = 6$ eV.

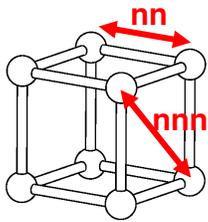 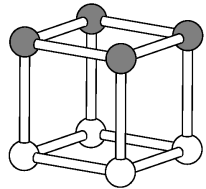 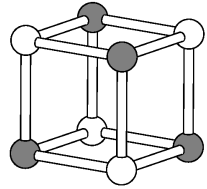

FM (0.0, 0.0)   AF1 (-5.0, 0.95)   AF2 (-5.7, -0.28)

…

4. S. Kamba, D. Nuzhnyy, P. Vaněk, M. Savinov, K. Knížek, Z. Shen, E. Šantavá, K. Maca, M. Sadowski, and J. Petzelt, Europhys. Lett. **80**, 27002 (2007).

5. V. Goian, S. Kamba, J. Hlinka, P. Vaněk, A. A. Belik, T. Kolodiazhnyi, and J. Petzelt, Eur. Phys, J. B **71**, 429 (2009).

6. K. A. Müller and H. Burkhard, Phys. Rev. B **19**, 3593 (1979).

7. T. R. McGuire, M. W. Shafer, R. J. Joenk, H. A. Halperin, and S. J. Pickart, J. Appl. Phys. **37**, 981 (1966).

8. K. A. Müller, Phys. Rev. Lett. **2**, 341 (1959).

9. P. A. Fleury, J. F. Scott, and J. M. Worlock, Phys. Rev. Lett. **21**, 16 (1968).

10. R. S. Krogstad and R. W. Moss, Bull. Am. Phys. Soc. **7**, 192 (1962).

11. E. J. Kirkpatrick, K. A. Müller, and R. S. Rubins, Phys. Rev. **135**, A86 (1964).

12. H. Unoki and T. Sakudo, J. Phys. Soc. Jpn. **23**, 546 (1967).

13. Th. Von Waldkirch, K. A. Müller, and W. Berlinger, Phys. Rev. B **7**, 1052 (1973).

14. G. Shirane and Y. Yamada, Phys. Rev. **177**, 858 (1969).

15. R. A. Cowley, W. J. L. Buyers, and G. Dolling, Solid St. Comm. **7**, 181 (1969).

16. J. M. Worlock, J. F. Scott, and P. A. Fleury, in "Light scattering analysis of Solids", ed. G. B. Wright (Springer, New York, 1969) p. 689.

17. A. Bussmann-Holder, H. Büttner, and A. R. Bishop, Phys. Rev. Lett. **99**, 167003 (2007).

18. A. Bussmann-Holder, J. Köhler, R. K. Kremer, and J. M. Law, Phys. Rev. B **83**, 212102 (2011).

19. T. R. McGuire, M. W. Shafer, R. J. Joenk, H. A. Alperin and S. J. Pickart, J. Appl. Phys. 37, 981 (1966).

20. C.-L. Chien, S. DeBenedetti and F. De. S. Barros, Phys. Rev. B 10, (1974).

21. E. H. Jacobsen and K. W. H. Stevens, Phys. Rev. **129**, 2036 (1963).

22. R. Migoni, H. Bilz, and D. Bäuerle, Phys. Rev. Lett. **37**, 1155 (1976).

23. H. Bilz, A. Bussmann, G. Benedek, H. Büttner, and D. Strauch, Ferroelectrics **25**, 339 (1980).

24. H. Bilz, H. Bilz, G. Benedek, and A. Bussmann-Holder, Phys. Rev. B **35**, 4840 (1987).

25. A. Bussmann-Holder and H. Büttner, Nature (London) **360**, 541 (1992).

26. R. O. Bell and G. Rupprecht, Phys. Rev. **129**, 90 (1963).

27. A. Migliori, J. L. Sarrao, W. M. Visscher, T. M. Bell, M. Lei, Z. Fisk, and R. G. Leisure, Physica (Amsterdam) **183B**, 1 (1993).

28. J. F. Scott, J. Bryson, M. A. Carpenter, J. Herrero-Albillos, and M. Itoh, Phys. Rev. Lett. **106**, 105502 (2011).

29. A. Bussmann-Holder, H. Beige, and G. Völkel, Phys. Rev. B **79**, 184111 (2009).

30. A. Bussmann-Holder, Phys. Rev. B **56**, 10762 (1997).

31. J. D. Axe, J. Harada, and G. Shirane, Phys. Rev. B **1**, 1227 (1970).

32. A. Koreeda, R. Takano, and S. Saikan, Phys. Rev. Lett. **99**, 265502 (2007).

33. (a) G. Kresse and J. Hafner, Phys. Rev. B **47**, 558 (1993); (b) G. Kresse and J. Furthmüller, J. Comput. Mater. Sci. **6**, 15 (1996); (c) G. Kresse and J. Furthmüller, Phys. Rev. B **54**, 11169 (1996).

34. J. P. Perdew, K. Burke, and M. Ernzerhof, Phys. Rev. Lett. **77**, 3865 (1996).

35. T. Katsufuji and H. Takagi, Phys. Rev. B **64**, 054415 (2001).

1636. S. L. Dudarev, G. A. Botton, S. Y. Savrasov, C. J. Humphreys, and A. P. Sutton, Phys. Rev. B **57**, 1505 (1998).

37. P. Blaha, K. Schwarz, G. K. H. Madsen, D. Kvasnicka and J. Luitz, WIEN2k (Vienna University of Technology, 2001) (ISBN 3-9501031-1-2).

38. M.-H. Whangbo, H.-J. Koo and D. Dai, *J. Solid State Chem.* 176, 417 (2003); D. Dai and M.-H. Whangbo, J. Chem. Phys. 114, 2887 (2001); D. Dai and M.-H. Whangbo, J. Chem. Phys. 118, 29 (2003).

39. H. J. Xiang, C. Lee and M.-H. Whangbo, Phys. Rev. B 76, 220411(2007); H.-J. Koo and M.-H. Whangbo, Inorg. Chem. 47, 128 (2008); H.-J. Koo and M.-H. Whangbo, Inorg. Chem. 47, 4779 (2008).

40. J. S. Smart, Effective Field Theory of Magnetism, Saunders, Philadelphia, 1966.